\documentclass{scrartcl}
\usepackage{amsmath}
\usepackage{slashed}

\def\be{\begin{equation}}
\def\ee{\end{equation}}
\def\rref#1{(\ref{#1})}
\def\scri{{\cal I}}
\def\vell{{\vec \ell}}
\def\p{\partial}

\title{Odd-dimensional de Sitter Space is Transparent}
\author{Philip Lagogiannis\thanks{philip.lagogiannis@mail.mcgill.ca},
  Alexander Maloney\thanks{maloney@physics.mcgill.ca}, \&
  Yi Wang \thanks{wangyi@hep.physics.mcgill.ca}
  \\
\\  \textit{Physics Department, McGill University, Montreal, H3A2T8, Canada}}
\date{}

\begin{document}

\maketitle

\thispagestyle{empty}

\begin{abstract}
We consider quantum field theory in de Sitter space, focusing on the
cases of scalars, spin 1/2 fields, and symmetric and anti-symmetric tensor fields of arbitrary spin.  The free field equations in global coordinates can be reduced to a one dimensional Schrodinger problem which possesses a remarkable structure; the potential is of an algebraically special type which appears as a multi-soliton solution of the KdV equation.  In an odd number of spacetime dimensions these potentials are ``transparent'' in the sense that their reflection coefficients vanish identically.  This has a remarkable consequence for physics in de Sitter space.  It means that odd dimensional de Sitter space is transparent in the sense that a quantum state with no particles at past infinity will evolve into a state with no particles in the far future.  This feature has been previously noted for scalar excitations, but the corresponding higher spin behaviour (and the proof using algebraic techniques) is new.
\end{abstract}

\newpage
\setcounter{page}{1}


\section{Introduction}

Space-time curvature leads to particle production.  This effect arises in the study of quantum field theory in curved geometries and has profound observational and theoretical consequences.   In an eternally inflating de Sitter geometry the particle production takes a particularly simple form; in the standard vacuum state an inertial observer observes a bath of thermal radiation emitted by the cosmological horizon \cite{GH}.  
During inflation the universe was approximately de Sitter, and the effects of this particle production have been observed indirectly in the cosmic microwave background. 
 
In this note we investigate free quantum field theory in de Sitter space and discover that this apparently staid and hoary subject still has a few surprises left in store.  We will describe a remarkable  structure arises when one studies de Sitter field theory in global coordinates.  In every case which we have studied, the field equations reduce to an effective one dimensional quantum mechanics problem which takes a special algebraic form.  For fields of integer spin the quantum mechanical potential is a multi-soliton solution to the KdV equation known as the Poschl-Teller potential.  When the number of space-time dimensions is odd, the potentials describe an integer number of coincident solitons.  This potential has the unusual property that it is reflectionless, i.e. the reflection coefficients vanish for all values of the frequency. For the case of half-integer spins the equations of motion  reduce to a similar one-dimensional Schrodinger problem which can be regarded as a complexified cousin of the Poschl-Teller potential.\footnote{In fact, precisely this potential also arises in a somewhat different context,  in the study of non-Hermitian Hamiltonians with PT symmetry.}  In each case the quantum spectrum can be solved using algebraic techniques related to supersymmetry.  When the number of space-time dimensions is even -- including the most important case of $D=4$ -- a similar structure arises.  The potentials can be thought of as describing a half-integer number of coincident KdV solitons and, while they are not reflectionless, can be solved using similar algebraic techniques based on supersymmetry.

The physical implications of this are remarkable.  In de Sitter space there are three natural sets of modes that one can use to define Fock space operators for free field theory.  There are those with no physical excitations at $\scri^+$, those with no excitations at $\scri^-$ and those which are obtained by analytic continuation from Euclidean signature.   Each of these choices form a complete basis of solutions to the wave equation.  Hence the modes are linearly related; the coefficients in the linear transformation between these different modes are known as Bogoliubov coefficients.  

The reflection and transmission coefficients of the auxilliary one dimensional quantum mechanics problem described above are precisely the Bogoliubov coefficients relating the modes at $\scri^-$ with those at $\scri^+$.  The reflectionless property implies that the mode solutions are, up to a phase factor, equal.  Thus, in an odd number of space-time dimensions de Sitter space is transparent, in the sense that a state with no particles in the far past will evolve into one with no particles in the far future.  At any intermediate time the state will contain particles; only when measured at future infinity do these excitations disappear.

In fact, this property has been noted previously in discussions of de Sitter invariant vacuum states referred to as Motolla-Allen or $\alpha$ vacua \cite{Mottola:1984ar, Allen:1985ux}.  Each choice of mode decomposition defines a different quantum vacuum state; the three sets of modes described above define the $|in\rangle$ (no particles at $\scri^-$), $|out\rangle$ (no particles at $\scri^+$) and $|E\rangle$ (Euclidean or Bunch-Davies) states, respectively.   It was observed in \cite{Bousso:2001mw} that for scalar fields $|in\rangle=|out\rangle$ in an odd number of space-time dimensions.  The goal of the present paper is to generalize this result to higher spin fields using the algebraic techniques described above.

Before proceeding, we note that the Euclidean vacuum is the only de Sitter invariant vacuum state with the same short-distance structure as the usual Minkowksi vacuum.  
If all modes are placed in the $|in\rangle$ or $|out\rangle$ vacuum then the resulting state is de Sitter invariant but has a non-standard short-distance structure which affects physics at length scales much shorter than the de Sitter radius.  This leads to problems when considering interacting field theory (see e.g. \cite{ Einhorn:2002nu, Kaloper:2002cs, Banks:2002nv}).  Nevertheless, the structure of the $|in\rangle$ and $|out\rangle$ wave functions still contains important physical information.  For example, while it seems unphysical to place arbitrarily short distance modes in the $|in\rangle$ and $|out\rangle$ states, it is certainly  reasonable to study states in which certain finite wavelength modes are placed in these states as opposed to the standard Euclidean vacuum.  Such states are essentially finite norm excitations of the Bunch-Davies vacuum and have been proposed as candidate states for quantum fields during inflation \cite{Einhorn:2003xb, deBoer:2004nd}.  

Moreover, the properties of the $|in\rangle$ and $|out\rangle$ wave functions may be relevant for quantum gravity in an eternally inflating universe.  For example, they (and other similar states) have appeared in various discussions of the dS/CFT correspondence  \cite{Strominger:2001pn} (see e.g. \cite{Anninos:2011jp} for a recent discussion of related issues).

In the present paper we will restrict ourselves to the study of formal properties of free wave equations in de Sitter space, leaving potential applications to cosmology for the future.  In the following section we describe in detail the reflectionless property of the scalar wave equation.  The physical content of this section has appeared in the literature before, but we will emphasize a somewhat different approach based on algebraic techniques.  In
section 3 the discussion is extended to Dirac fields, before turning
to higher spin fields in section 4. Section 5 contains a brief 
discussion of open questions.

\section{Scalar Fields in de Sitter Space}

\subsection{Wave Equations in Global Coordinates}

In this section we consider a free scalar field $\phi$ in $n$ dimensional de Sitter space. In global coordinates the metric is
\begin{align}
  {ds^2\over \ell^2}=-d\tau^2+\cosh^2\tau ~ d\Omega_{n-1}^2~
\end{align}
where $d\Omega_{n-1}^2$ is the metric on the sphere $S^{n-1}$.
This space-time is the maximally symmetric solution
of Einstein's equations with a positive cosmological constant $\Lambda=(n-2)(n-1)/(2\ell^2)$.  
We henceforth use units where the de Sitter radius is $\ell=1$.  In these coordinates asymptotic past/future infinity are given by $\tau\to\pm\infty$.

It was noticed in \cite{Bousso:2001mw} that when $n$ is odd there is a sense in which
there is no production of quantum excitations of $\phi$ due to the curvature of de Sitter space.  
In this section, we will reproduce this result and emphasize
the connection with the theory of reflectionless potentials.  The physical content of this section is not particularly new and we suspect that
the appearance of KdV solitons in the scalar wave equation has been noticed  by many people
over the years (see e.g. \cite{polyakov}).  There is a vast literature on the subject of quantum field theory in de Sitter space; we refer the reader to \cite{BD} for an introduction.

To quantize the field $\phi$, we start by choosing a complete basis of solutions to the wave equation
\begin{align}
  (\nabla^2-m^2) \phi = 0 ~,
\end{align}
where $m$ is the mass of the field.  In this paper we will, for simplicity, focus on massive fields in order to avoid the various subtleties which arise for massless fields in de Sitter space.  
 
We denote by $\{\phi_{\vec \ell}\}$ a complete basis of solutions to the wave equation, where 
${\vec \ell}$ is a set of parameters labelling the modes.  We expand $\phi$ in terms of raising and
lowering operators as
$$
\phi = \sum_{\vec \ell} a_\vell \phi_{\vec \ell} + {a_\vell}^\dagger \phi_\vell^*
$$ 
In order to guarantee that 
$a_\vell$ and $a_\vell^\dagger$ obey the usual algebra of raising and lowering operators, we must require the $\phi_\vell$ to be normalized with respect to the standard Klein-Gordon inner product
$$
\langle f | g\rangle = i \int_\Sigma d\Sigma^\mu (f^* \p\mu g - \p_\mu f^* g)
$$
Here the integral is over an arbitrary spacelike slice $\Sigma$ through spacetime, which we can take to be a sphere $S^{n-1}$ at constant time.
 Associated to the mode decomposition $\phi_{\vec \ell}$ we can define a vacuum state, which is the state in the Fock space annihilated by $a_\vell$ for all values of $\vec \ell$.

In curved space-time there is no canonical choice of mode decomposition $\phi_{\vec \ell}$, and hence no canonical choice of vacuum state.
One natural choice of modes are those $\phi_\vell$ which are purely positive frequency at $\tau\to -\infty$.  These modes 
will oscillate like $e^{-i \omega \tau}$ at $\tau\to-\infty$, and define a state $|in\rangle$.  From the point of view of a local observer at $\scri^-$ this state will appear to have no particles.  Likewise, 
the modes which go like $e^{-i\omega\tau}$ at $\tau \to +\infty$ define a state $|out\rangle$ with no particles in the 
far future.  Neither of these states are equivalent to the Euclidean vacuum $|E\rangle$ which is defined using the Hartle-Hawking construction.  This vacuum is constructed using modes which are smooth on the lower half of sphere $S^{n}$ obtained by Wick rotation of the de Sitter metric to Euclidean signature.
 The Euclidean modes have both positive and negative frequency parts at $\scri^\pm$.  

It is instructive to work out the form of the modes explicitly. 
In global coordinates, the de Sitter laplacian is
\begin{align}
  \nabla^2=-(\cosh\tau)^{1-n}\partial_\tau\left[(\cosh\tau)^{n-1}\partial_\tau\right]
  +(\cosh\tau)^{-2}\Delta_{S^{n-1}}~,
\end{align}
where $\Delta_{S^{n-1}}$ is the Laplacian on $S^{n-1}$.  
We can then expand these wave functions in terms of spherical harmonics on $S^{n-1}$,
which are unit normalized and obey
$$
\Delta_{S^{n-1}}Y_\vell=-l(l+n-2)Y_\vell.
$$
Here the index $\vell = (l,m_1,m_2\dots)$ is a collective index which labels the various spherical harmonics, and $l$ is a non-negative integer.
The full mode solution can be expanded as 
\begin{align}\label{modes}
  \phi_\vell=
  u_\vell(\tau)~(\cosh\tau)^{-(n-1)/2} ~Y_{l,m_1,m_2,\ldots}(\Omega_{n-1})
\end{align}
where $\Omega_{n-1}$ is a point on the sphere.
We have chosen the normalization factor ${\cosh^{(n-1)/2}\tau}$ in $\phi_\vell$ for future convenience.  With this choice,  the 
Klein-Gordon normalization condition is simply
\be\label{norm}
u{\p_\tau u}^* - u^*\p_\tau u=i
\ee

The equation of motion then takes the form
\begin{align}\label{equivQM}
\left(  \partial_\tau^2 +
    \left(\frac{2l+n-3}{2}\right)\left(\frac{2l+n-1}{2}\right)
    \frac{1}{\cosh^2\tau}
    +\left(m^2-\frac{(n-1)^2}{4}\right)
  \right) u=0~.
\end{align}
This is the time-independent Schrodinger equation for a one dimensional quantum mechanical system, where $u(\tau)$ is regarded as the wave function and $\tau$ as the position variable.  The Klein-Gordon normalization condition \rref{norm} is just the statement that the probability current of the wave function $u(\tau)$ is conserved.\footnote{With a different choice of prefactor in \rref{modes} we would obtain a different equation for the probability current, along with a friction term proportional to $\p_\tau u$ in the equation of motion \rref{equivQM}.}  The  potential appearing in this effective Schrodinger equation is known as the Poschl-Teller potential \cite{pt}
\be\label{pt}
V = - {L(L+1)\over \cosh^2 \tau},~~~~~L={2l+n-3\over 2}
\ee
We note that the parameter $L$ is an integer when the number of space-time dimensions is odd.

The potential \rref{pt} has several truly remarkable properties when $L$ is an integer.
First, its reflection coefficients vanish identically.  Second, it possesses exactly $L$ normalizable bound states.  
 These two results can be derived in one of three ways.  The first is by direct solution of the wave equation in terms of Legendre or hypergeometric functions.  The second uses algebraic techniques related to supersymmetry.     The third (which is overkill in the present situation) uses inverse scattering techniques and relies on the fact that the potential \rref{pt} is the solution of the KdV equation  describing L coincident solitons.  We review these results in the next subsection, where we use algebraic techniques.
  
However, before giving a mathematical proof of the reflectionless properties of the potential \rref{pt},  we note that in the present context this property has a simple physical interpretation.  As $\tau\to\pm\infty$ the solutions to the wave equation (\ref{equivQM}) are plane waves\footnote{For simplicity we focus here on the case of massive fields $m^2>\frac{(n-1)^2}{4}$ so that $\omega$ is real.  Thus we are considering massive fields whose Compton wavelength is smaller than the Hubble scale.}
$$
u \sim e^{\pm i \omega \tau}~~~~~\omega = \sqrt{m^2 - \frac{(n-1)^2}{4}  }
$$
The modes which define the $|in\rangle$ and $|out\rangle$ state behaves as $e^{-i \omega \tau}$ at $\tau \to -\infty$ and $\tau \to +\infty$, respectively.  The reflectionless property is simply the statement that the $|in\rangle$ modes $u^{in}$ have the following behaviour at $\tau \to \pm \infty$: 
$$
u^{in} \sim \left\{{e^{-i \omega \tau}~~~~~~~~\tau\to -\infty \atop T(\omega) e^{-i \omega \tau}~~~~~\tau\to\infty} \right.
$$
where $T(\omega)$ is a phase which  will be computed in the next section.   Thus the state which is purely positive frequency in the far past remains purely positive frequency in the far future.  This means that  $|in\rangle$ and $|out\rangle$ modes are related by
$$
u^{in} = T(\omega) u^{out}
$$
and the $|in\rangle$ and $|out\rangle$ vacua are equal.  
Thus the scalar wave equation in odd-dimensional de Sitter space is transparent.

\subsection{Algebraic Approach to Reflectionless Potentials}

We now wish to demonstrate that potential \rref{pt} is reflectionless.  
In fact \rref{pt} is a member of a remarkable family of quantum mechanical potentials which have the property that any incoming wave will be transmitted through with unit
probability, regardless of the frequency of the incoming wave.  These are known variously as reflectionless or transparent potentials, and are sometimes discussed in textbooks on elementary quantum mechanics (see e.g. \cite{LL}).  We will describe here a slightly more sophisticated approach based on supersymmetry, which allows for an elegant algebraic 
computation of the spectra and scattering data of certain reflectionless potentials.  The advantage of this technique is that it allows us to demonstrate the reflectionless property, and to compute the phase $T(\omega)$, without needing to solve any differential equations. Our discussion here is not new and can be found in any number of references; we refer the reader to \cite{SQM}  for a  pedagogical treatment.

We start by considering a Hamiltonian 
\be
H = p^2 + V~~~~~p=-i{\p \over \p\tau}
\ee
where $V(\tau)$ approaches a constant at $\tau\to\pm \infty$.  We would like to ask under what circumstances the scattering solutions 
\be
H \psi = \omega^2 \psi
\ee
are reflectionless, in the sense that 
\be
\psi \sim \left\{{e^{i \omega \tau}~~~~\tau \to - \infty \atop T(\omega) e^{i \omega \tau}~~~~~\tau \to +\infty}\right.
\ee
for all values of the frequency $\omega$.

At first sight one might guess that the only potential whose reflection coefficients vanish identically is the trivial
potential $V= {\rm constant}$.  To see why this is not the case, it is useful to consider the following pair of one dimensional Hamiltonians
$$
H_+= A_+A_-~,\qquad H_-=A_- A_+~
$$
where we have defined the two operators
$$
A_\pm = p \pm i W(\tau)
$$
The function $W(\tau)$, referred to as a superpotential, will be assumed to approach a constant as $\tau\to \pm \infty$. 
Writing out the two Hamiltonians $H_\pm$ explicitly
$$
H_\pm = p^2 + V_\pm ~,\qquad V_\pm = W^2 \mp W'~.
$$
We see that the potentials $V_\pm$ approach constants as  $\tau\to\pm\infty$.  The spectra of $H_\pm$ will include scattering states as well as possible bound states. 

In fact, for each eigenstate $|\psi\rangle$ of $H_\pm$ one can construct a corresponding eigenstate of $H_\mp$ with the same energy by acting with the operator $A_\mp$:
$$H_\pm|\psi\rangle = E |\psi\rangle ~~~~\implies~~~~H_\mp \left(A_\mp |\psi\rangle \right) = E \left(A_\mp |\psi\rangle\right)$$
Thus the spectra of $H_+$ and $H_-$ are nearly identical.  The only exception is if $H_\pm$ has a zero energy state $|\psi\rangle$ which is annihilated by $A_\mp$. In this case there is no corresponding non-trivial eigenstate of $H_\mp$.

This implies that the reflection and transmission coefficients of $H_\pm$ are related in a simple manner. 
To see this, consider a scattering solution of $H_+$, i.e. an eigenfunction $\psi(\tau)$ of $H_+$ obeying
$$
\psi\sim\left\{{e^{i\omega\tau} + R_+(\omega) e^{-i\omega\tau},~~~~~\tau\to-\infty\atop T_+(\omega) e^{i \omega \tau},~~~~~~~~~~~~~ \tau\to \infty}\right.
$$ 
where $R_+(\omega)$ and $T_+(\omega)$ are the reflection and transmission coefficients of $H_+$.
 The operator $A_-$ which converts this into a eigenfunction of $H_-$ takes a simple form at $\tau\to\pm\infty$ where $W(\tau)$ is constant, allowing us to determine the scattering coefficients of $H_-$:
 $$
 R_-(\omega) = \frac{\omega+i W(-\infty)}{-\omega+i W(-\infty)} R_+(\omega)~,\qquad T_-=\frac{\omega-i W(\infty)}{ \omega-i W(-\infty)} T_+(\omega)~.
 $$ 
 
This construction can be described in terms of supersymmetry.  To see this, note that the matrix operators 
 $$
 H = \left(
 \begin{array}{cc}
   H_-&0 \\0&H_+ 
 \end{array}\right)~, \qquad
Q_+ = \left( 
 \begin{array}{cc}
   0&0 \\A_+&0 
 \end{array}\right)~, \qquad
Q_- = \left(
 \begin{array}{cc}
   0&A_- \\0&0 
 \end{array}\right)~, \qquad
 $$
 obey the usual supersymmetry commutation relations
 $$
 [Q_\pm, H]=0~,\qquad \{Q_+,Q_-\}=H~.
 $$
For this reason, the Hamiltonians $H_\pm$ and potentials $V_\pm$ will be referred to as superpartners.
Indeed, the spectral degeneracy described above is nothing more than the usual supersymmetric relation between even and odd spin states.  Of course, as we are working with quantum mechanics rather than quantum field theory, the use of the word ``spin" here does not represent the transformation property of the states under rotation, but simply reflects a grading of the Hilbert space into even and odd parts.  States which are annihilated by $A_\pm$ are the analogues of BPS states, and the mismatch described above between the zero energy spectra of $H_\pm$ is the statement that the index $Tr (-1)^F$ may be non-zero.
 
It is now straightforward to construct reflectionless potentials.
We start by seeking the potential which is the superpartner of the constant potential.  That is, we seek a superpotential $W(\tau)$ for which $V_-=1$\footnote{Here we have restricted our attention to case where the potential is a positive constant, which can be set equal to one by rescaling $\tau$.  If we replace the right hand side of this equation by a non-positive constant we will obtain another solution which is formally reflectionless, but which diverges at finite $\tau$ so is less interesting for our purposes.}
$$
W^2 + W' = 1
$$
The solution is
$$
W = \tanh \tau
$$
and the corresponding superpartner potential is
\be\label{v2}
V = 1 - {2\over \cosh^2  {\tau}}~.
\ee
We therefore conclude that this potential is reflectionless.  Note that if we shift $V$ by a constant the potential will remain reflectionless, so we can freely omit the constant term.

We can now generate a second reflectionless potential by identifying this $V$ as the $V_-$ of another superpotential and then solving for the corresponding $V_+$.  Iterating this procedure will yield an infinite series of reflectionless potentials of the form
\be
V_L = - {L(L+1) \over \cosh^2 \tau}
\ee
for any integer value of $L$.   This is proven by noting that the potentials $V_{L}$ and $V_{L-1}$ are superpartners connected by the superpotential
\be
W_L = L \tanh \tau~.
\ee
In this computation, as above, we have omitted constant terms in $V_L$ which do not affect the reflectionless property.
The potentials $V_L$ described above are the famous Poschl-Teller potentials, whose reflectionless properties have been known since the early days of quantum mechanics.

Using the algebraic technique described above it is straightforward to obtain expressions for the scattering states along with the transmission coefficients $T(\omega)$.  Given a solution $\psi_{L-1}(\tau)$ of the potential $V_{L-1}$, a solution of $V_L$ is obtained by applying the operator $A_L = p + i W_L$.  So starting with the plane wave solution $e^{i\omega\tau}$ of the constant potential $V_0$ we obtain the scattering solution of the potential $V_L$
\be\label{psiL}
\psi_L = A_L \dots A_1 e^{i\omega\tau}~.
\ee
Since the $A_L$ take a simple form at $\tau\to \infty$ we can easily extract the transmission coefficients
$$
T(\omega) = \prod_{n=1}^L {\omega+in \over \omega-in}~.
$$
This is pure phase, as it must be since the reflection coefficient vanishes.


A similar argument can be used to show that the potential $V_L$ has $L$ bound states.  Indeed, this is reflected in the fact that the 
transmission coefficient has $L$ poles at the complex momenta $\omega=i,2i,\dots,Li$.  When $L$ is not integer one can still use algebraic methods to simplify the construction of scattering and bound states, even though the potential will not be reflectionless.  For example, when $L$ is half integer (the case of interest for even dimensional de Sitter space) the scattering states are given by acting with $A_L \dots A_{3/2}$ on the basic scattering solutions of the $L=1/2$ potential.

Finally, we note that the algebraic procedure described above does not give the most general form of a reflectionless potential.  More general examples can be constructed using inverse scattering techniques; we refer the reader to the literature for a discussion of these techniques (see for example  \cite{Drazin:1989qi}). 

\section{Spinor Fields in de Sitter Space}

In this section we consider free spin $1/2$ fields in de Sitter space. We will show that, as in the scalar case, odd
dimensional de Sitter spaces is transparent.  Rather than solving the Dirac equation explicitly, we will simply show that the 
corresponding effective potential which appears in global coordinates is reflectionless, following the argument of the previous section.

\subsection{Wave Equation in Global Coordinates}

The Dirac equation is
\begin{align}\label{DDirac}
  (\Gamma^\mu D_\mu+m)\Psi = 0 ~.
\end{align}
Our conventions for Gamma matrices and spin connections are given in Appendix A.  We will now write out the Dirac equation 
in global coordinates. 

We start by considering the case where the number of spacetime
dimensions $n$ is odd, with $n\geq 3$.  In this
case a Dirac spinor in $n$ dimensions has the same number of components as a Dirac spinor in $n-1$ dimensions.  
It is then just a matter of working out the components of the spin connection.  In global coordinates, the Dirac equation is \begin{align}
  \left(\partial_\tau+\frac{n-1}{2}\tanh\tau\right)
 \gamma_0
  \Psi
  + \frac{\tilde{\slashed\nabla}\Psi}{\cosh\tau} 
  + m\Psi = 0 ~.
\end{align}
where $\tilde{\slashed\nabla}$ is the Dirac operator on $S^{n-1}$. 

We now expand in terms of a complete basis of eigenfunctions of the Dirac operator on $S^{n-1}$.
We will follow the conventions of Camporesi and Higuchi \cite{Camporesi:1995fb}, who defined a complete orthonormal basis of spinors ${\hat \chi}^{(\pm)}_{lm}$ on the sphere. They obey the eigenvalue equation
\begin{align}
  -i\gamma^0 \tilde{\slashed\nabla}\hat\chi_{lm}^{(\pm)} 
  = \pm \frac{2l+n-1}{2} \hat\chi_{lm}^{(\pm)} ~,~~~~~
-i \gamma^0
  \hat\chi_{lm}^{(\pm)}=\hat\chi_{lm}^{(\mp)}~.
\end{align}
Here $l$ and $m$ are integers. 

We now expand 
\begin{align}
    \Psi = \sum
       f_{lm}^{(\pm)}(\tau) (\cosh\tau)^{-(n-1)/2}~ \hat\chi_{lm}^{(\pm)}(\Omega^{n-1})~, 
\end{align}
where $f_{lm}^\pm$ are some functions of $\tau$ and we have introduced a factor of $(\cosh\tau)^{-(n-1)/2}$ for future convenience.
The Dirac equation then takes the form\footnote{As in the scalar case, our prefactor in has been chosen so that there is no term linear in $\p_\tau$.}
\begin{align}\label{dirac}
  \left(\partial_\tau^2 
    \pm\frac{iL \sinh\tau}{\cosh^2\tau} 
    +\frac{L^2}{\cosh^2\tau} + m^2 
  \right) f_{lm}^{(\pm)} =0 ~.
\end{align}
where $L$ is defined by
\begin{align}\label{Lis}
  L=\frac{(2l+n-1)}{2}
\end{align}
We note that $L$ is an integer, as $n$ is odd.
For reasons that will become clear below, this potential should be regarded as a complex version of the Poschl-Teller potential.

It is straightforward to show that in even dimensions the Dirac equation reduces to the same one dimensional equation \rref{dirac}, with $L$ again given by \rref{Lis}.
We start by recalling that in this case an $n$ dimensional Dirac spinor has twice the number of components as an $n-1$ dimensional
spinor.
So writing out the Dirac equation in global coordinates, we find
\begin{align}
  \left\{
    \left(\partial_\tau + \frac{n-1}{2}\tanh\tau\right)
    \left(
    \begin{array}{cc}
      { 0}&{\bf 1} \\{\bf 1}&{ 0} 
    \end{array}
  \right)
  +
  \frac{1}{\cosh\tau}
  \left(
    \begin{array}{cc}
      0& \tilde{\slashed\nabla} \\ -\tilde{\slashed\nabla}&0 
    \end{array}
  \right)
  -im
\right\} \Psi = 0 ~,
\end{align}
where $\tilde{\slashed\nabla}$ is the Dirac operator for $S^{n-1}$.  Following \cite{Camporesi:1995fb}, we use the complete basis of spinors $\chi_{lm}^{(\pm)}$ on $S^{n-1}$ which obey
\begin{align}
  \tilde{\slashed\nabla} \chi_{lm}^{(\pm)}(\Omega^{n-1}) = \pm \frac{2l+n-1}{2} \chi_{lm}^{(\pm)}(\Omega^{n-1})~.
\end{align}
We then expand in terms of these fields 
\be
\Psi=\sum \left(f_{lm}^{(\pm)} \atop g_{lm}^{(\pm)} \right)
(\cosh\tau)^{-(n-1)/2} \chi_{lm}^{(\pm)}~,
\ee
where $f_{lm}^{(\pm)}$ and $g_{lm}^{(\pm)}$  are functions of $\tau$.
The Dirac equation reduces then reduces to equation \rref{dirac}, along with a similar equation for $g_{lm}^{(\pm)}$.  We note that in the even dimensional case the parameter $L$ appearing in the equation of motion is half-integer rather than an integer.

The wave equation \rref{dirac} is reflectionless when $L$ is an integer, just as in the scalar case.  We give a simple proof of its reflectionless property in the following subsection using algebraic methods.\footnote{The reflectionless property of this potential has been noted in a different context by \cite{Levai:2001zs, CDV,Ahmed:2005zz}.}
This implies  that, as in the scalar case, odd dimensional de Sitter space is transparent for Dirac fields.  In the even dimensional case there will be particle production, as the parameter $L$ is now half-integer.

Finally, it is worth noting that the case of two dimensional de Sitter space is rather special, since in this case there are two possible spin structures.  They correspond to taking the fermions to have either periodic or antiperiodic boundary conditions as we go around the spatial circle.  The effective potential described above corresponds to the anti-periodic case, which is the natural spin structure which generalizes to the higher dimensional spheres.  With periodic boundary conditions, however, the parameter $l$ appearing in \rref{Lis} is half-integer.  Therefore  $L$ is an integer and the potential is reflectionless.  Thus with the appropriate choice of fermion boundary conditions two dimensional de Sitter space is transparent as well.

\subsection{Complexified Reflectionless Potentials}

We will now generalize the discussion of section 2.2 to show that the equation of motion \rref{dirac} is reflectionless, just as in the scalar case.  

We first note that if equation \rref{dirac} is regarded a one dimensional Schrodinger equation then the corresponding  potential is complex and Hamiltonian is non-Hermitian.  This  does not reflect any underlying sickness in our theory; the Dirac equation \rref{DDirac} describes Hermitian time evolution as usual.  It is only means that, when written in terms of an auxilliary one dimensional potential, the Dirac equation can be formulated in terms of a  non-Hermitian Hamiltonian.\footnote{As an aside we note that while this auxilliary system is non-Hermitian it is PT symmetric in the sense of \cite{Bender:1998gh}.}  We note that this potential approaches a constant at $\tau\to\pm\infty$, so one can compute its scattering solutions just as in section 2.2.  It is these scattering solutions which are reflectionless.

The proof of this is a straightforward generalization of the discussion in section 2.2.  In that case $W$ was taken to be real, which is the usual case of interest in quantum mechanics.  The operators $A_+$ and $A_-$ were Hermitian conjugates of one another and the $H_\pm$ were Hermitian.  However, when $W$ is allowed to be complex an additional family of complex reflectionless potentials can be constructed. The potentials are of the form
\be
V_L = - {L^2 \over  \cosh^2 \tau} + {i L } {\sinh \tau \over \cosh^2 \tau}
\ee
for integer values of $L$.  
To show that these potentials are reflectionless, we note that (up to constant terms which may be dropped) the potentials $V_L$ and $V_{L-1}$ are superpartners, connected by the complex superpotential
\be
W_L= (L-\frac{1}{2})\tanh \tau + \frac{i}{2 \cosh \tau}~.
\ee
When $L=0$ this gives the trivial constant potential; thus the whole family of potentials is reflectionless for integer values of $L$.

As before, we can now obtain expressions for the scattering states along with the transmission coefficients $T(\omega)$.  Defining $A_L = p + i W_L$ the scattering solutions are again
\be\label{psiL}
\psi_L = A_L \dots A_1 e^{i\omega\tau}~.
\ee
The transmission coefficients are
$$
T(\omega) = \prod_{n=1}^L {\omega+i(n-1/2) \over \omega-i(n-1/2)}~.
$$

\section{Higher Spin Fields in de Sitter space}

We will now generalize the previous discussion to free fields of arbitrary integer spin, focusing on the case of differential form fields and symmetric tensor fields.  In both cases the wave equations will be reflectionless in an odd number of space-time dimensions.

\subsection{Differential Form Fields}

We first consider the case of massive form fields of arbitrary rank.  

The action of a free $p$-form field $A_p$ is
\begin{align}
  S=\int d^nx \sqrt{-g} 
  \left[
    -\frac{1}{2(p+1)!}F_{\lambda_1\cdots\lambda_{p+1}}F^{\lambda_1\cdots\lambda_{p+1}}
    -\frac{m^2}{2p!}A_{\lambda_1\cdots\lambda_p}A^{\lambda_1\cdots\lambda_p}
  \right]~.
\end{align}
where
\begin{align}
  F_{\lambda_1\cdots\lambda_{p+1}} 
  = (p+1)\partial_{[\lambda_1}A_{\lambda_2\cdots\lambda_{p+1}]} ~.
\end{align}
The equations of motion are
\begin{align}\label{formEOM}
  (p+1)\partial_{\lambda_1}
  \left\{
    \sqrt{-g}~g^{\lambda_1\rho_1}\cdots g^{\lambda_{p+1}\rho_{p+1}}
    \partial_{[\rho_1}A_{\rho_2\cdots\rho_{p+1}]}
  \right\}
  -m^2 \sqrt{-g} A^{\lambda_2\cdots\lambda_{p+1}}=0~.
\end{align}
Note that we may act with $\partial_{\lambda_2}$ on equation (\ref{formEOM}) to obtain the  constraint equation
\begin{align}\label{formConstraint}
  \partial_{\lambda_1}\left(\sqrt{-g}A^{\lambda_1\cdots\lambda_p}\right)=0~.
\end{align}

Our strategy will be to divide the $p$-form field $A_p$ into two components: a component $A^{i_1\dots i_{p}}$ with indices in the $S^{n-1}$ directions and a component $A^{0i_1\dots i_{p-1}}$ with one time-like index.  Roman indices $i_1, \dots$ denote sphere directions.

Let us start with the components $A^{0i_1\dots i_{p-1}}$, which may be regarded as a $p-1$ form on the sphere.  Setting $\lambda_p=0$ in equation (\ref{formConstraint})  
\begin{align}
  \partial_{i_2}\left( \sqrt{-g}A^{0i_2\cdots i_p} \right) =0~.
\end{align}
we see that $A^{0i_2\cdots i_p}$ is a co-closed $(p-1)$-form on sphere.
On the other hand, setting $\lambda_a\neq 0$, $a=2,\ldots, p$,
equation (\ref{formConstraint}) becomes 
\begin{align}\label{constraint1}
  \sqrt{-g}\left[\partial_\tau A^{0i_2\cdots i_p}
  +(n-1)\tanh\tau A^{0i_2\cdots i_p}\right]
  +\partial_{i_1}\left( \sqrt{-g}A^{i_1\cdots
      i_p}\right)
  =0~.
\end{align}
Using equation (\ref{constraint1}), the $\lambda_2=0$ part of
equation (\ref{formEOM}) is
\begin{align}
  \left[
    \partial_\tau^2 + (n-1+2p)\tanh\tau~\partial_\tau+2p(n-1)+m^2
    -\frac{(2p-1)(n-1)}{\cosh^2\tau}-\frac{\tilde\Delta}{\cosh^2\tau}
  \right]A^{0i_2\cdots i_p}=0~,
\end{align}
where $\tilde\Delta=d\delta+\delta d$ is the  Hodge-de Rham Laplacian on $S^{n-1}$. 
In order to write this in Schrodinger form we first introduce the $p-1$ form $f^{i_2\cdots i_{p-1}}$
defined by
\begin{align}
  A^{0i_2\cdots i_p} = (\cosh\tau)^{-(n-1+2p)/2} f^{i_2\cdots i_{p}}~.
\end{align}
We then expand in a basis of eigenforms of the Laplacian on the sphere, which were described in \cite{CH}.  The operator $\tilde\Delta$ acting on co-closed $(p-1)$-forms has eigenvalues $-(l+p-1)(l+n-p-1)$ where $l$ is a non-negative integer.  Expanding in this basis, a component of the form field $f$ will obey the Schrodinger equation
\begin{align}
  \partial_\tau^2 f +
  \left\{
    \frac{L(L+1)}{\cosh^2\tau}
    +\left[m^2-\frac{1}{4}(n-1-2p)^2\right]
  \right\}f=0~.
\end{align}
where
\be
L = \frac{2l+n-3}{2}
\ee
Up to a shift of mass, this equation is identical to the scalar case
(\ref{equivQM}).  In particular, it is reflectionless when $n$ is odd since in that case  $L$ is an integer. 

We now consider the component $A^{i_1\cdots i_p}$, which can be regarded as a $p$-form living on the sphere.  We will denote this component ${\hat A}$, and denote by $d$ and $\delta $ the differential and codifferential on $S^{n-1}$.    To show that $\hat A$ is reflectionless, we first note that $\delta \hat A$ is completely fixed by  the constraint equation (\ref{constraint1}), so can be expressed in terms of  $A^{0i_2\cdots i_p}$.  Therefore $\delta \hat A$ is reflectionless.  
Likewise, acting with $d$ on (\ref{formEOM})  we obtain
\begin{align}
  \left[\partial_\tau^2+(n-1-2p)\tanh\tau\partial_\tau 
    -\frac{\tilde \Delta }{\cosh^2\tau}+m^2\right](d \hat A) =0~.
\end{align}
We then remove the $\partial_\tau$ term by defining
\begin{align}
  d\hat A = (\cosh\tau)^{-(n-1-2p)/2}f~
\end{align}
Using the fact that the Laplacian $\tilde \Delta $ has eigenvalues $-(l+p)(l+n-p-2)$ the equation becomes
\begin{align}\label{ddeltabeta}
  \partial_\tau^2 f+
  \left\{
    \frac{L(L+1)}{\cosh^2\tau}
    +\left[m^2-\frac{1}{4}(n-2p-1)^2\right]
  \right\}f=0~~~~~~L=\frac{2l+n-3}{2}.
\end{align}
Again, the equation is identical to (\ref{equivQM}) and is reflectionless in an odd number of space-time dimensions when $L$ is an integer. 
We have now shown that both $d\hat A$ and $\delta \hat A$ are reflectionless.  It follows that $\tilde \Delta \hat A$ is reflectionless.  As the sphere has no non-trivial cohomology, there are no non-trivial harmonic forms on the sphere and the operator $\tilde \Delta$ is invertible.  Thus $\hat A$ is reflectionless. 

We conclude that all of the components of a differential form field are reflectionless in odd dimensional de Sitter space.

\subsection{Symmetric Tensor Fields}

We now consider a massive symmetric tensor field $h_{\mu_1\dots \mu_r}$ of rank $r$ which  obeys the equation of motion
\be  
(\nabla^2-m^2) h_{\mu_1\mu_2\cdots\mu_r}=0~.
\ee
and is divergenceless and traceless  
\begin{align}
  g^{\mu_1\mu_2} h_{\mu_1\mu_2\cdots\mu_r}=0~,~~~~~  \nabla^{\mu_1}h_{\mu_1\mu_2\cdots\mu_r}=0~,
\end{align}
As in the previous cases, the reflectionless behaviour can be derived in a straightforward way from the wave equation.  We will proceed by relating the equations to those for symmetric tensor harmonics on the sphere described by \cite{Higuchi:1986wu}.

To write out the wave equation, we start by noting that 
for any $p$ between $0$ and $r$, the component $h_{\tau\cdots\tau i_1\cdots i_{p}}$ can be regarded as a symmetric tensor of rank $p$ on the sphere $S^{n-1}$. 
We can then define the $p^{th}$ component of $h$ as 
\begin{align}\label{hp}
  h_{\tau\cdots\tau i_1\cdots i_{p}} = 
  (\cosh\tau)^{2p-r}f_{i_1\cdots i_p}~
\end{align}
where the $\tau$ dependent factor has been introduced for future
convenience.  We will abbreviate $f_{i_1\dots i_p}$ by $f_p$.  It is
then straightforward (but tedious) to write out the wave equation in
global coordinates.  We find
\begin{align}
    \left[
    \partial_\tau^2+(n-1)\tanh\tau\partial_\tau
    -\frac{\tilde\nabla^2-p}{\cosh^2\tau}+m^2
  \right]f_p &
  \nonumber\\
  - 2ip \frac{\sinh\tau}{\cosh^3\tau}\tilde\nabla f_{p-1}&
  +p(p-1)\frac{\sinh^2\tau}{\cosh^4\tau}\eta f_{p-2}=0 ~,
  \label{eq:wave-tensor}
\end{align}
Here $\eta$ and $\tilde\nabla$ denote the metric and covariant derivative
on the sphere.  We have not written out the indices explicitly; $\eta f_{p-2}$ and $\tilde \nabla f_{p-1}$ are the rank $p$ tensors on the sphere which are symmetrized over their indices.
The primary subtlety is that the wave equation (\ref{eq:wave-tensor}) couples components \rref{hp} with different values of $p$.

To solve this equation, we note that for every solution $h_{\mu_1\dots
  \mu_r}$ there is a value of $p$ (possibly zero) such that
$h_{\tau\cdots\tau  i_1\cdots i_{q}}$ vanishes for all $q<p$.  We can
then expand $f_p$ in terms of a basis of symmetric tensor spherical
harmonics on $S^{n-1}$ as
\begin{align}
  f_p = \sum_{l, m} u^{(p)}_l(\tau) Y^{(p)}_{l m}(\Omega_{n-1})~.
\end{align}
This basis of spherical harmonics was described explicitly in \cite{Higuchi:1986wu}.  Here $l, m$ are integers.
These tensors obey the eigenvalue equation
\be
\tilde\nabla^2 Y^{(p)}_{l m} = [-l(l+n-2)+p] Y^{(p)}_{l m} ~.
\ee
Then the equation of motion for $u_l^{(p)}$ is 
\begin{align}
  &
  \left[
    \partial_\tau^2+(n-1)\tanh\tau\partial_\tau
    +\frac{l(l+n-2)}{\cosh^2\tau}+m^2
  \right]u_l^{(p)} =0 ~,
  \label{eq:wave-tensor-lowest}
\end{align}
From this it is easy to see that $u_l^{(p)} (\cosh\tau)^{(n-1)/2}$ satisfies the Poschl-Teller 
equation (\ref{equivQM}). Thus $f_p$ (and hence $h_{\tau\cdots\tau i_1\cdots i_{p}}$) is reflectionless in odd dimensions.

The equations for the higher rank components $f_{p+1}$, $\ldots$,
$f_r$ are more
complicated and involve couplings between the fields.  However the
wave equation \rref{eq:wave-tensor} allows us to solve for these
higher rank components in terms of $f_p$.  The computation is
essentially identitical to that in \cite{Higuchi:1986wu} so we will
just summarize the answer.  We find that
%
 \begin{align} \label{eq:f-T}
   f_{p+q}=
    c_0^{(q)} T_{p+q}^{(q)}+\cdots+
    (-1)^k c_k^{(q)} \eta^k
    T^{(q-2k)}_{p+q-2k}+\cdots
\end{align}
solves the equation of motion (\ref{eq:wave-tensor}),
where $\eta_{ij}$ is metric on the sphere and the $T$'s are functions on
$S^{n-1}$, defined as
\begin{align}
  T^{(q)}_{p+q}  = \sum_{k=0}^{[q/2]} \alpha(n,k,p,q) \eta^k
  \tilde\nabla^{q-2k} Y_{p}~,
\end{align}
where $[q/2]$ is the integer part of $q/2$. Here $\alpha(k,n,p,q,l)$ is a
number, recursively defined in \cite{Higuchi:1986wu}. Again all
indices are symmetrized. 

The coefficients $c_0^{(q)}$ in equation (\ref{eq:f-T}) vanish for $q<0$.  For $q=0$ we have
\begin{align}
 c^{(0)}_{0}= u_{l}^{(p)}(\tau)~,
\end{align}
The wave equation, along with the divergenceless and traceless conditions, then reduces to the following equation
\begin{align}
  c^{(q)}_k = \frac{1}{2^k
    k!}\frac{(p+q)!}{(p+q-2k)!}
  \left[\prod_{a=1}^{k}\frac{1}{n+2(p+q-2k+a-1)}\right]\frac{c_0^{(q-2k)}}{(\cosh\tau)^{2k}}~,
\end{align}
\begin{align}
  &
  \frac{(q+1)(n+2p+q-3)
    \left[
      (p+q)(p+q+n-2)-l(l+n-2)
    \right]
  }{(p+q+1)(n+2p+2q-3)} c_0^{(q+1)}
  \nonumber\\
  & - \frac{(p+q)c_0^{(q-1)}}{(n+2p+2q-3)\cosh^2\tau}
  +i\left[\partial_\tau +(n+p+q-2)\tanh\tau\right] c_0^{(q)}=0
\end{align}
This allows us to solve recursively for $c_k^{(q)}$ in terms of $u_{l}^{(p)}$. As
$u_{l}^{(p)}$ is reflectionless for odd $n$, all of these coefficients will similarly be 
reflectionless. 

We conclude that massive symmetric tensor fields are
reflectionless in odd dimensional de Sitter space.
 
\section{Conclusion and Discussion}

We have learned that the presence of curvature does not necessarily lead to particle production in cosmological settings.  
This observation has relied on the relationship between free wave equations in de Sitter space in global coordinates and the theory of reflectionless potentials.  This relationship has largely escaped notice in the literature because most work has focused on wave equations in the Poincare patch where the spatial slices are flat.   

In the present paper we have considered various types of massive free fields and given proofs of transparency on a case by case basis. It is possible that there are other types of fields which do not have this reflectionless property; this would be nice to investigate.  For example, it would be interesting to study massless or very light (Compton wavelength of order Hubble scale) fields in de Sitter space, which exhibit well known subtleties at the free and interacting level.  To this end it may be useful to rephrase the results presented in this paper in a more unified manner based on the representation theory of the de Sitter group.  

We conclude by mentioning three important open questions.  The first is whether other cosmologies are also transparent in the sense described in this paper.  The answer to this question is almost certainly yes, as the classes of potentials described above are but the simplest examples of reflectionless potentials. The second question is whether this transparency remains once interactions are present.  In this case the answer is not clear; in the absence of a deeper symmetry explanation for the present results we see no reason why the transparency property should be preserved.  The final, and most important, question is whether these results are of observational relevance for theories of the early universe.  We leave this as a challenge for the future.

\section*{Acknowledgments}
We thank A. Adams, D. Anninos, R. Bousso, A. Higuchi, D. Marolf, I. Morrison, P. Horava, L. Leblond and E. Silverstein
for discussions.  We thank the participants of the McGill workshop
``Infrared effects in de Sitter Space" for encouraging the authors to
write up these results.  A. M. thanks his loyal undergraduate researchers M. Emelin,  N. Lvov and P. Sabella-Garnier for stimulating discussions.  This work is supported by the
Fonds Qu\'eb\'ecois de la Recherche sur la Nature et les Technologies, the
Natural Sciences and Engineering Research Council of Canada and the
Institute of Particle Physics.

\appendix

\section{Conventions on Gamma Matrices}

We follow the conventions of 
\cite{Camporesi:1995fb} which we summarize here.  
Lorentzian  gamma matrices obeying
$$
 \{\gamma^a, \gamma^b\} = 2 \eta^{ab}~ .
$$  
are used to define curved space Gamma matrices 
$
\Gamma^\mu = e^\mu_a \gamma^a
$. The Dirac operator $D_\mu = \p_\mu + \Omega_\mu$ is defined in terms of the spin connection
\begin{align}
  \Omega_\mu=\frac{1}{2}w_{\mu ab}\Sigma^{ab}~, \quad
  \Sigma^{ab}=\frac{1}{4}[\gamma^a,\gamma^b]~,\quad 
  w_{\mu ab}=e^{\nu}_a\nabla_\mu e_{\nu b}~.
\end{align}
Euclidean signature gamma matrices are defined recursively and given in even dimensions by
\begin{align}
  \gamma_E^n=
    \left(
    \begin{array}{cc}
      { 0}&{\bf 1} \\ {\bf 1}& { 0} 
    \end{array}
  \right) ~, 
  \qquad
  \gamma_E^j=
    \left(
    \begin{array}{cc}
      0&{i \tilde\gamma_E^j} \\ {-i \tilde\gamma_E^j}&0 
    \end{array}
  \right) ~,
\end{align}
where $j=1, \ldots, n-1$ and $\tilde\gamma_E^j$ are Euclidean gamma matrices in one dimension less.
In odd dimensions
\begin{align}
  \gamma_E^n =
    \left(
    \begin{array}{cc}
      {\bf 1}&{0} \\ {0}&{\bf -1} 
    \end{array}
  \right) ~, 
  \qquad
  \gamma_E^j = \tilde\gamma_E^j ~,
\end{align}
where $j=1, \ldots, n-1$ and $\tilde\gamma_E^j$. 
The Gamma matrices used in section 3 are Minkowski signature, and related to the Euclidean Gamma matrices by
\begin{align}
  \gamma^0=i\gamma_E^n ~, \qquad \gamma^j=\gamma_E^j ~.
\end{align}

\end{document}